\documentclass[aps,prxquantum,reprint,twocolumn,superscriptaddress]{revtex4-2}
\usepackage{amsmath}
\usepackage{amssymb}
\usepackage{graphicx}
\usepackage{booktabs,graphicx,rotating}
\usepackage{amsmath}
\usepackage{amssymb}
\usepackage{graphicx}
\usepackage{bbm}
\usepackage{xcolor}
\usepackage[colorlinks=true,urlcolor=blue,menucolor=blue,citecolor=blue,linkcolor=blue]{hyperref}
\usepackage{todonotes}

\newcommand{\ket}[1]{|{#1}\rangle}

\makeatletter
\def\myfnsymbol#1{\expandafter\@alph2\csname c@#1\endcsname}
\def\@myfnsymbol#1{\ensuremath{\ifcase#1\or \dagger\or \ddagger\or *\or
   \mathsection\or \mathparagraph\or \|\or **\or \dagger\dagger
   \or \ddagger\ddagger \else\@ctrerr\fi}}
\let\@fnsymbol\@myfnsymbol
\makeatother

\newtheorem{theorem}{Theorem}[section]
\newtheorem{corollary}{Corollary}[theorem]
\newtheorem{lemma}[theorem]{Lemma}

\begin{document}

\author{Violeta N. Ivanova-Rohling}
\email{violeta.ivanova-rohling@uni-konstanz.de}
\affiliation{Department of Physics, University of Konstanz, D-78457 Konstanz, Germany}
\affiliation{Zukunftskolleg, University of Konstanz, D-78457 Konstanz, Germany}

\author{Niklas Rohling}
\email{niklas.rohling@uni-konstanz.de}
\affiliation{Department of Physics, University of Konstanz, D-78457 Konstanz, Germany}
\author{Guido Burkard}
\email{guido.burkard@uni-konstanz.de}
\affiliation{Department of Physics, University of Konstanz, D-78457 Konstanz, Germany}

\title{Discovery of an exchange-only gate sequence for CNOT with record-low gate time using reinforcement learning}
\begin{abstract}
    Exchange-only quantum computation is a version of spin-based quantum computation that entirely avoids the difficulty of controlling individual spins by a magnetic field and instead functions by sequences of exchange pulses.
    The challenge for exchange-only quantum computation is to find short sequences that generate the required logical quantum gates. A reduction of the total gate time of such synthesized quantum gates can help to minimize the effects of decoherence and control errors during the gate operation and thus increase the total gate fidelity.
    We apply reinforcement learning to the optimization of exchange-gate sequences realizing the CNOT and CZ two-qubit gates which lend themselves to the construction of universal gate sets for quantum computation.
    We obtain a significant improvement regarding the total gate time compared to previously known results.
\end{abstract}

\maketitle

\section{Introduction}

Quantum computing has been a strongly growing field in the last years due to its potential to solve certain problems efficiently that are hard on a classical computer.
The growth of the field is driven by advances in gate fidelity and the number of qubits for scalable quantum computing platforms. Among those platforms are superconducting qubits \cite{arute2019quantum,kjaergaard2020superconducting}, Rydberg atoms \cite{evered2023high}, trapped ions \cite{bruzewicz2019trapped}, and spin qubits in semiconductor quantum dots \cite{burkard2021semiconductor}.
Specifically, spin qubits are promising with respect to scaling due to their small size and synergy with silicon-based technology and due to recent advances in gate fidelity \cite{xue2022quantum,noiri2022,madzik2022}.
In the original spin-qubit setting \cite{loss1998quantum}, 
each qubit is represented by the spin of an electron or hole trapped in a semiconductor quantum dot. 
Computations in such single-electron or single-hole spin qubits are based on controlling the exchange interaction between the spins and the local time-dependent magnetic fields acting on individual spins.
A logical two-qubit universal gate, such as the controlled-NOT (CNOT) gate, is implemented by using exchange-interaction-based SWAP$^\alpha$ operations controlled by inter-dot voltage combined with magnetic-field-controlled single-qubit gates.
These physical SWAP$^\alpha$  gates are the result of the exchange operation, where $\alpha$ is the normalized time parameter for which the exchange interaction is pulsed and the SWAP gate is switched on.
This means that $\alpha$ denotes the gate time in units of $\pi/J$ where $J$ is the strength of the exchange interactions when switched on.
One of the challenges for quantum computing based on spins in quantum dots is the single-spin control that necessitates the modulation of a strongly non-homogeneous magnetic field on short time scales or the realization of a strongly inhomogeneous magnetic field using on-chip micro-magnets \cite{pioro2008}.
%
The necessity for this is completely avoided in an alternative approach which encodes one logical qubit using three physical spins.
For this encoding, the exchange interaction is sufficient to implement universal quantum gates  \cite{divincenzo2000universal,bacon2000universal}, and thus the control of the local magnetic field is no longer necessary.
This paradigm of quantum computation is referred to as spin \emph{exchange-only} computation and has been subject to great experimental advances recently \cite{weinstein2023universal}.

Various approaches to exchange-only computation exist, described in \cite{russ2017three}.  This physical platform has since been theoretically and experimentally investigated, and a large number of practical implementations of quantum dot systems for three-spin qubits have been developed, for more detail refer to \cite{burkard2021semiconductor}.  

We will consider the exchange-only computational model, described in  Ref.~\cite{divincenzo2000universal}, where each qubit is encoded using three physical spins (spin-$\frac{1}{2}$ particles), and where one- and two-qubit quantum gates on the logical qubits are implemented by sequences of SWAP$^\alpha$  gates (switching on and off the exchange interaction between pairs of spin particles) applied to the physical qubits. The exchange interaction can be completely switched off by a sufficiently large voltage barrier between the quantum dots (then the gate is off), and only through pulsing the voltage, the exchange interaction is switched on. 

\begin{figure}
\includegraphics[width=\columnwidth]{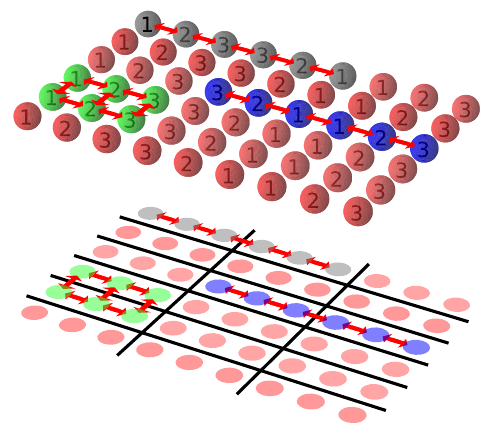}
\caption{A possible two-dimensional arrangement of physical spins
where each rectangle with three spins (labeled 1,2,3) defines one logical qubit.
The definition of the logical qubit depends on the order of the physical qubit because each logical qubit, $a\ket{0}_l + b\ket{1}_l$ is defined for the total-spin-0 subspace as
$\ket{0}_l=(\ket{010}-\ket{100})/\sqrt{2}$, $\ket{1}_l=(2\ket{001}-\ket{010}-\ket{100})/\sqrt{6}$, for the total-spin-1 subspace, see \cite{fong2011universal}.
Here, $|ijk\rangle=|i\rangle_1 |j\rangle_2 |k\rangle_3$ where the numbers 1,2, and 3, correspond to the labels in the figure.
We clearly see that the definition depends on the order of the physical qubits (spins), specifically the spins labeled 1 and 2 host a singlet in state $\ket{0}_l$ and a triplet in state $\ket{1}_l$.
Further, note that the coupling of neighboring logical qubits depends on the arrangement and on the order of the physical qubits (within one row or within neighboring rows).
This results in three distinct coupling scenarios between pairs of logical qubits (highlighted in gray, blue, and green) which we label ``33" (gray), ``11" (blue), and 2D (green) which are all considered in this paper.
The exchange couplings that can be used in each scenario are indicated with red arrows.}
\label{fig:grid}
\end{figure}

The cost of disposing of single-spin rotations in exchange-only quantum computation is twofold, (1)  the necessity of an extended physical system, i.e., a larger number of physical spins to represent the qubit register, and (2) logical quantum gates that need to be synthesized by a sequence of several physical exchange operations, rather than a single application of exchange in the standard spin qubit paradigm. More specifically, in the case of two-qubit gates, several applications of the exchange interaction between at least five pairs of spins are involved in controlling the system of two logical qubits, see Fig.~\ref{fig:grid}. A vital aspect for exchange-only computation to be practically relevant is thus to optimize the efficiency of the gates needed for quantum computation. This is where this paper provides a novel method that is shown to allow for a substantial improvement by applying reinforcement learning (RL).
%
Efficiency can be looked at in several terms: minimizing the number of pulses, time steps, or total time necessary.
In this paper, we will focus on minimizing the total gate time for a fixed value $J$ of the exchange coupling when switched on.
We will consider both, the case of exchange gates applied in parallel when possible and exchange gates applied sequentially which can be advantageous for avoiding cross-talk \cite{weinstein2023universal}.
Minimizing the time needed for a desired gate is beneficial with respect to gate fidelity as noise acts on the system for a shorter time while the gate sequence is performed.
For pulses of fixed duration with varied exchange strength, as in \cite{weinstein2023universal}, the actual time needed for the sequence will depend only on the number of pulses (sequential) or number of time steps for pulses applied in parallel.
However, we note that there is a lower limit to the gate time set by the maximum available value of $J$.
Additionally, minimizing the total time normalized to a fixed $J$ \footnote{The quantity we refer to as normalized sequential time is termed \textit{exchange angle} in \cite{weinstein2023universal}.} then corresponds to operating at smaller exchange coupling which can reduce charge noise \cite{weinstein2023universal}.

Moreover, when arranging the physical qubits in a two-dimensional square lattice, different connections between neighboring logical qubits are present, see Fig.~\ref{fig:grid}. Each of these different arrangements yields a distinct optimization problem.
In quantum computing, a CNOT gate is universal when combined with single-qubit gates, which makes finding (efficient) exchange-only sequences to realize the CNOT gate an important problem.  In \cite{divincenzo2000universal}, the first exchange-only universal gate set consisting of single-qubit rotations and a CNOT was presented. 
In \cite{hsieh2003explicit}, an exact specification of a universal logical gate-set using four spins to encode a single qubit was presented. The authors use extensive numerical optimization in order to obtain an optimized CNOT gate sequence with 27 parallel nearest-neighbor exchange interactions or 50 serial gates.

\begin{figure}
\begin{center}
\includegraphics[width=\columnwidth]{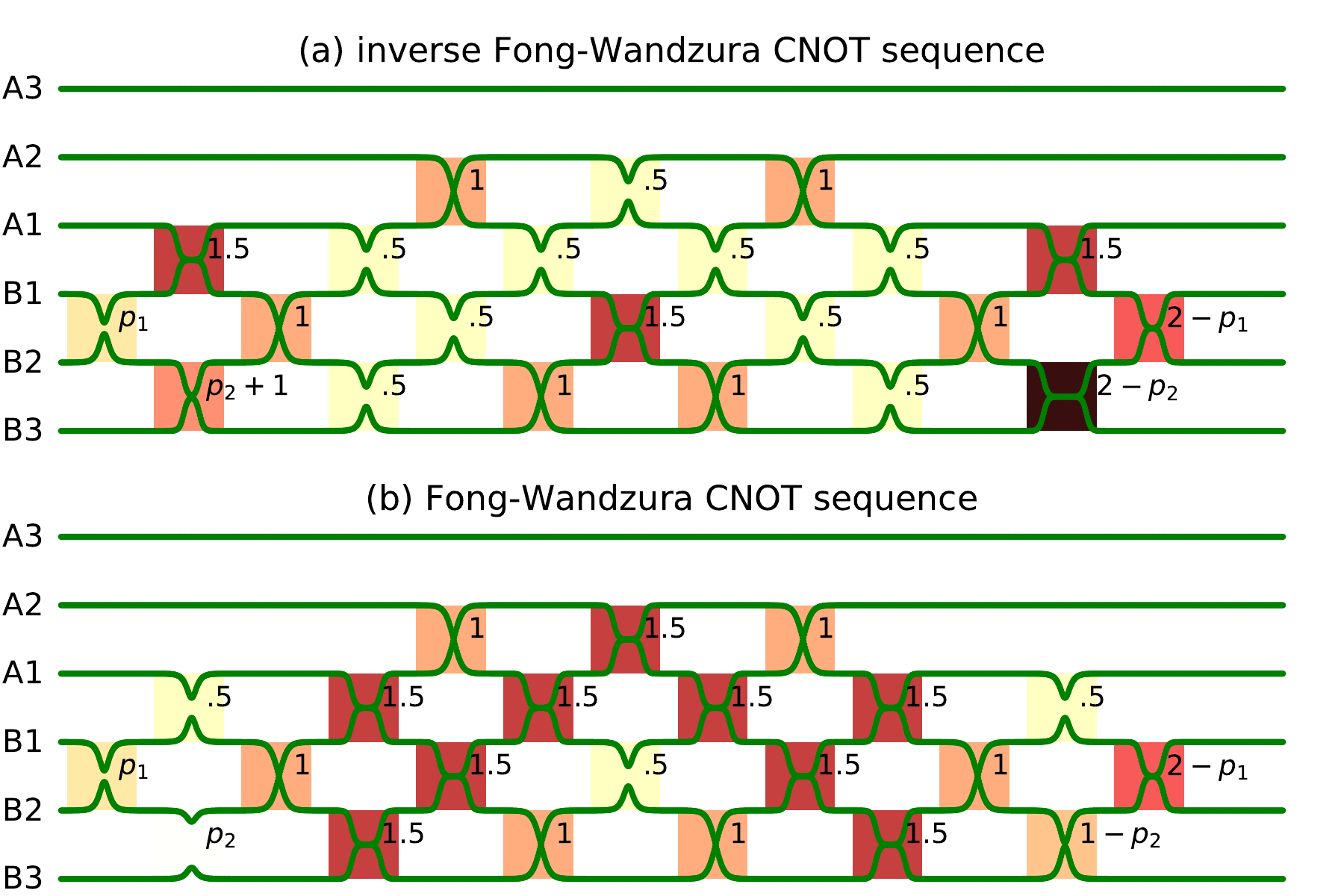}
\end{center}
\caption{ (a) Inverse FW sequence compared to (b) the FW sequence from Ref.~\cite{fong2011universal}. Both sequences require twenty-two pulse and 13 time steps, while their parallel times are $T_p=13.89\,\pi/J$ and $T_p=15.89\,\pi/J$, respectively, and their sequential times are $T_s=20$ and $T_s=24$, respectively.
Logical qubits $A$  and $B$  are arranged as
  shown.  Numbers after $A$ and $B$ label the physical
  qubits. The  SWAP powers $\alpha$ are displayed explicitly in the gates.  $\alpha=1$
  corresponds to a full SWAP operation, up to a global phase. Here, we have introduced 
  $p_1=\arccos(-1/\sqrt{3})/\pi$ and $p_2=\arcsin(1/3)/\pi$.
  The representation of the SWAP$^\alpha$ gates in this figure is inspired by the representation in \cite{weinstein2023universal}.}
\label{fig:CNOT}
\end{figure}

Different approaches have been utilized to find optimized sequences numerically \cite{divincenzo2000universal, hsieh2003explicit, fong2011universal}. 
The 
sequence for a CNOT  found by Fong and Wandzura, via the use of genetic algorithms \cite{fong2011universal}, see Fig.~\ref{fig:CNOT}, is currently the most efficient exact CNOT sequence known when the physical qubits are connected via nearest-neighbor interactions and they are in a linear chain architecture, see the area in Fig.~\ref{fig:grid} highlighted in blue. Importantly, despite the fact that this solution has been discovered numerically, it has a precise analytical description.
%
In \cite{setiawan2014robust}, gate sequences were found for logical two-qubit gate locally equivalent to CNOT for various connectivities by applying exhaustive search under the condition that all exchange gates are $\sqrt{\mathrm{SWAP}}$ or products thereof.
Aside from a large number of purely numerical approaches, it has been possible to come up with an analytic derivation of the optimal Fong-Wandzura (FW) CNOT sequence  \cite{zeuch2016simple}. 
%
%
Furthermore, analytical considerations with regards to \emph{leakage} were utilized in combination with numerics \cite{van2019approximate} to simplify the search problem and construct another set of gate sequences realizing the CNOT gate. Under certain assumptions,  the solution presented in Ref.~\cite{van2019approximate} is more efficient than the FW sequence if one considers total time as the efficiency criterion.  
Other efficient universal two-qubit gates have also been investigated,
such as a gate locally equivalent to the CPHASE gate  \cite{zeuch2020efficient} that is potentially valuable in the currently available NISQ quantum devices.
Leakage errors in exchange-only spin qubits can be approached by a reset-if-leaked procedure and, via numerical optimization, by a leakage correcting gate sequence  \cite{langrock2020reset}. 
 
Numerous advances in implementing quantum dot systems for three-spin qubits have been made 
\cite{laird2010coherent, gaudreau2012coherent, medford2013self, medford2013quantum, kim2014quantum, eng2015isotopically, reed2016reduced, cao2016tunable, thorgrimsson2017extending}.
Recently, Weinstein \textit{et al.}~\cite{weinstein2023universal} presented a two-qubit exchange-only system implemented using an array of six $^{28}$Si/SiGe quantum dots to achieve universal gates of very high operational fidelity. The fidelity of universal control of two encoded qubits was evaluated to be  $96.3\% \pm 0.7\%$ for encoded CNOT operations, and even higher  $(99.3\%\pm 0.5\%)$ for encoded SWAP, demonstrating substantial progress towards achieving fault tolerance and computational acceleration with this approach. 


The problem of optimizing gate sequences for two-qubit logical gates is high-dimensional. In our work, we use an intelligent optimization \cite{pham2012intelligent} approach enhanced by an RL algorithm, suitable for continuous search spaces.
This allows us to explore a vastly larger search space by enforcing much fewer assumptions on the optimization problem in comparison to \cite{setiawan2014robust}.
We aim to optimize the total time of the exchange-only gate sequences representing exact CNOT and exact CZ gates with varying connection topologies and 
find gate sequences for all three arrangements shown in Fig.~\ref{fig:grid}. 
For the linear ``11" arrangement highlighted in blue in Fig.~\ref{fig:grid}, we find gate sequences representing CNOT gates.
Notably one of the sequences we find, presented in Fig.~\ref{fig:CNOT} (a), has a shorter total time than the original FW sequence and the RL approach found it from scratch.
We discuss the relation to other known gate sequences in Sec.~\ref{sec:discussion}.


Importantly, we demonstrate the usefulness of a reinforcement-learning-based approach for optimizing exchange-only sequences, which can be seamlessly extended to optimize different universal gates,  and gate sequences with different architectures and different types of exchange interactions by simply redefining the cost function. 
The main aspects of the reinforcement learning approach we use for gate sequence optimization are visualized in Figure \ref{fig:RLalgorithm}, and full details of the approach are given in Sec.~\ref{sec:methods}, as well as in Appendix~\ref{app:rldetails}.
\begin{figure*}
\includegraphics[width=15cm]{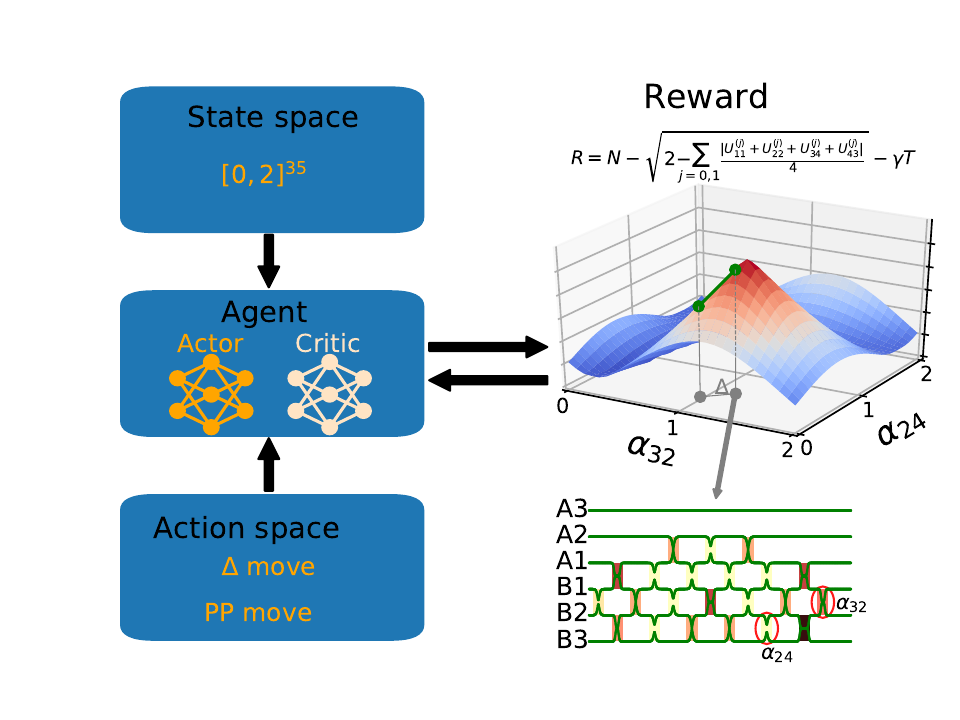}
\caption{The reinforcement learning (RL) algorithm used for CNOT gate sequence optimization.
A deep deterministic policy gradient (DDPG) actor-critic algorithm was used as an agent.
The state space is defined as 35 normalized time parameters $\alpha_i\in[0,2]$ of SWAP$^{\alpha_i}$ gates, operating on six spins $A_k$ and $B_k$ ($k=1,2,3$) representing two logical qubits $A$ and $B$, and arranged in the five-brick structure. Each state corresponds to a sequence of exchange pulses with a length of up to 14 time steps using parallelism.
The number of pulses can be smaller than 35 when some of the $\alpha_i$ are zero.
The action space is defined as two possible actions -- a $\Delta$ step, that potentially changes all parameters $\alpha_i$, and a PP-move, i.e., a partial Powell local optimization step, where a local optimization with a fixed number of iterations is performed. The number of iterations is small ranging from 0 (no local optimization is performed) to 12. 
The reward is based on the distance to the CNOT gate used by Fong and Wandzura \cite{fong2011universal} and the total time, which can be either the time $T=T_p$ for applying the SWAP$^{\alpha_i}$ gates in parallel where possible, or the time $T=T_s$ for applying the SWAP$^{\alpha_i}$ gates sequentially.}
\label{fig:RLalgorithm}
\end{figure*}

Additionally, we apply the RL algorithm to find optimized CZ gate sequences, see Sec.~\ref{app:resultsCZ}, CNOT sequences for a linear arrangement with the singlet-triplet qubit part on the edges (``33" arrangement, highlighted in gray in Fig.~\ref{fig:grid}) and obtain a sequence beating the one actually implemented in \cite{weinstein2023universal} with respect to sequential total gate time, see Sec.~\ref{sec:Weinstein} and Appendix \ref{app:resultsWeinstein} for details.
Furthermore, we search for optimized gate sequences for CNOT 
gates in the 2D arrangement of spins, highlighted in green in Fig.~\ref{fig:grid}, where seven pairs of spins can be coupled by exchange interactions, see Sec.~\ref{sec:2D} and Appendix \ref{app:2D}.

\section{Methods}
\label{sec:methods}
\subsection{Reinforcement learning for optimization problems}
Reinforcement learning is a class of machine learning algorithms, where an agent interacts with an environment and gets back a reward based on its actions. The goal of the agent is to learn a behavior that optimizes the total reward obtained.
 RL that uses neural networks as agents to learn the optimal policy is referred to as \emph{deep RL}. Recently, RL, and especially deep RL have been used with great success for numerous problems in various areas of physics, in general, \cite{martin2021reinforcement},  as well as quantum computing \cite{krenn2023}, in particular.
 More recently, RL has been used to learn appropriate optimizers that solve difficult optimization problems, or to \emph{learn to optimize},
examples include \cite{l2obenchmark,learningsparse,li2017learning,l20scalable, pmlr-v70-chen17e,compoptrl}.

The RL approaches to optimization show advantages over automating and accelerating the optimization of complicated problems. Instead of manually crafting classical optimizers, one can parameterize and learn optimization rules in a data-driven fashion. 

Yet another application of RL for optimization is to use the RL agent as a hybrid aspect of the optimizer to automatically guide the behavior of the optimizer in an intelligent way, suitable in particular for the problem at hand. This does not involve ``learning" to optimize on a similar task prior to the optimization task, but using the machinery of RL, and the stored experiences during the optimization procedure (the experience replay \cite{fedus2020revisiting}), to select the appropriate next steps in the optimization search. Based on the agent's prior experience and obtained reward, the next optimization behaviors are selected, instead of encoded, such as selecting exploration vs. exploitation behaviors, or parameter values. Examples include \cite{seyyedabbasi2023reinforcement,seyyedabbasi2021hybrid}, where different global optimization heuristics were combined with a simple Q-learning approach to intelligently choose between exploitation or exploration behavior of the heuristic, as well as intelligently set other parameters of the optimization heuristics. These intelligent optimization approaches were tested on known hard mathematical functions as benchmarks and were found to outperform other state-of-the-art methods that were not enhanced by RL. In \cite{karimi2022machine}, a review of hybrid approaches for optimization, that use RL as well as metaheuristics for combinatorial optimization is presented. In \cite{samma2016new}, a memetic particle swarm optimizer that uses RL to control optimization operations, related to choosing local search behavior and particle selection, was introduced. The method turned out to be successful on several benchmark optimization problems.
In this work, we use a deep RL approach to intelligently guide an optimizer to better optimize a gate sequence. The RL agent, based on previous experience, recorded in an experience buffer, and on previous rewards from the environment, predicts the optimality of an action. In this case, the action is a behavior of the optimizer.

\subsection{The five-brick structure}
\begin{figure}
\includegraphics[width=5cm]{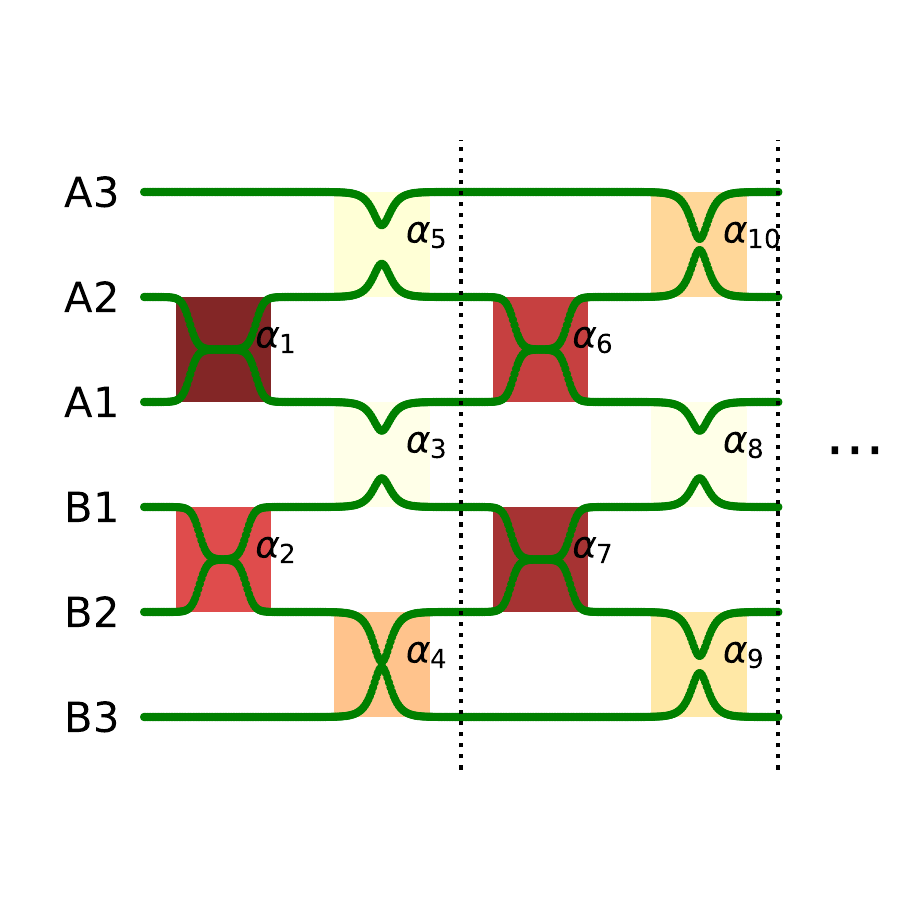}
\caption{The five-brick structure used to define the observation space of the RL algorithm.}
\label{fig:FB}
\end{figure}

In the search for the reset-if-leaked sequence \cite{langrock2020reset}, a brick-like structure of repeated patterns of physical exchange gates (SWAP-$\alpha$) was used. The brick structure is taking advantage 
of the commutation relation between the exchange interactions between the qubits in different subsystems.
Two exchange interaction operators commute if the exchange interaction is applied to pairs of logical qubits that do not share any common spins. Then the gate sequence is invariant under the interchange of the order of these operators. 
Here, since we are not trying to reproduce the FW gate sequence, but aim to improve it, we loosen the four-brick pattern structure to a general five-brick structure (Fig.~\ref{fig:FB}) that allows all six physical qubits to be affected by the sequence, which in principle enables a generalized search for other, potentially better sequences. 

\subsection{Reward function for the CNOT-gate}
In order to assess how well a gate sequence approximates a logical CNOT, the distance from CNOT is measured using the FW distance function introduced in \cite{fong2011universal},
\begin{equation}
\begin{split}
    d_{\rm FW}(U(\boldsymbol\alpha)) =& \left[2-\frac{1}{4}\left|U^{(0)}_{11} + U^{(0)}_{22} + U^{(0)}_{34} + U^{(0)}_{43}\right|\right.\\
               & \left.-\frac{1}{4}\left|U^{(1)}_{11} + U^{(1)}_{22} + U^{(1)}_{34} + U^{(1)}_{43}\right| \right]^{1/2},
\end{split}
\end{equation}
where $U^{(0/1)}$ is the $\boldsymbol\alpha$-dependent unitary matrix describing the overall gate sequence on the subspace for total spin zero or total spin one, respectively.
Here, $\boldsymbol\alpha=(\alpha_1,\alpha_2,\ldots)$ represents the list of exchange time parameters.
The function $d_{\rm FW}$ is a distance measure between a unitary matrix and the desired CNOT gate, taking advantage of the fact that CNOT as the target gate comprises only ones and zeros as matrix elements in the computational basis.
Furthermore, note that while the unitarity of $U^{(0/1)}$ is used, $d_{\rm FW}$ allows for different phase factors in the spin-0 and in the spin-1 subspaces.
The reward is  given by
\begin{equation}
    R(\boldsymbol\alpha) = N -d_{\rm FW}(U) -\gamma T,
    \label{eq:reward}
\end{equation}
where $T$ denotes the total gate time.
The last term rewards minimizing the total time needed for the gate sequence.
$N$ is a large number added for technical reasons, as negative rewards do not perform well.

\subsection{Reinforcement learning for gate sequence optimization}
\label{rlalgo}
Reinforcement learning (RL) can use the memory hash that is built from the learning experience in order to achieve intelligent optimization. The reward feedback, provided from the environment in the RL setting can improve the optimizer's behavior, and instead of choosing parameters of the optimization heuristics manually, the RL machinery can be used to guide the optimizer parameters in high-reward areas, with the actor-critic used to learn to predict the behavior of the optimizer that will optimize the reward. 

A visual representation of our RL approach is shown in Figure \ref{fig:RLalgorithm}, where the observation space consists of the possible values for the normalized times $\alpha_i$ for gate sequences of fixed length $35$, the action space consists of two types of actions, a small change of the normalized times of the sequence, and a partial local optimization (the derivative-free Powell's method) of a fixed small number of iterations. The RL agent learns the best way to optimize the total time of the sequence in an actor-critic approach, where both the actor and the critic are neural networks. The reward obtained by the agent at each step is based on the sequence distance to the exact CNOT or CZ gate and the total time of the gate sequence. 
To optimize the gate sequence, given the difficulty of the problem, we utilize RL to learn strategies to optimize the sequence, instead of manually selecting and parametrizing an optimizer.   We use the deep deterministic policy gradient (DDPG) \cite{lillicrap2015continuous} algorithm, which is an actor-critic algorithm \cite{konda1999actor} for RL with continuous state space, where the gate sequence is assumed to be constructed by a sequence of repeating the five-brick structure of a fixed length of 35 pulses, and the state space consists of the values of the normalized times $\alpha_i$ of the different SWAP$^{\alpha_i}$ gates, $i = 0,\ldots,34$. We use hybrid control, namely the action space has both continuous as well as discrete components. The continuous components are values that change the normalized times $\alpha_i$ at a given step, while the discrete component determines the number of iterations of a partial derivative-free optimization (partial use of Powell's method, \cite{fletcher1963rapidly}). The number of possible iterations can be 0, which allows for the case where no derivative-free partial optimization takes place, and only the values of the normalized times are varied. By \textit{partial optimization} here we mean that we fix the number of optimization iterations without the necessity of a local optimum to be achieved.  The goal is to learn a sequence of parameter values (starting points of the partial Powell algorithm) that result in the best gate sequence. As a reward we use a function based on the FW  measure for the distance from CNOT combined with the total time, see Eq.~(\ref{eq:reward}).


\subsection{Optimizing $T_s$ for CNOT with linear ``33" arrangement}
\label{sec:Weinstein}
We investigate the performance of our approach for optimizing the CNOT gate sequence, imposing the same constraints on qubit arrangement as in \cite{weinstein2023universal} in order to be able to compare the resulting gate sequences to the one used in Weinstein \textit{et al.}~\cite{weinstein2023universal}.
We enforce connectivity constraints of the physical qubits so that the singlet-triplet part of the logical qubit is on the outside of the gate sequence chain.
This yields the order of the linear-chain arrangement highlighted in gray in Fig.~\ref{fig:grid} which is A1, A2, A3, B3, B2, B1 (the ``33" arrangement).
This allows us to compare solutions discovered by our approach to the solution expressed in \cite{weinstein2023universal}, where such additional requirement was imposed.
We present the results in Sec.~\ref{sec:results}.

\subsection{2D Connecting topology}
\label{sec:2D}
In addition to the linear arrangement, we also consider the case where
the three physical qubits of one logical qubit are coupled to their counterparts of the other logical qubit, see the qubits highlighted in green in Fig.~\ref{fig:grid}. This requires an adjustment of the grouping of the exchange gates and complicates the optimization procedure, see Sec.~\ref{sec:results} for the results.

\section{Results}
\label{sec:results}
\begin{figure*}
\includegraphics[width=\textwidth]{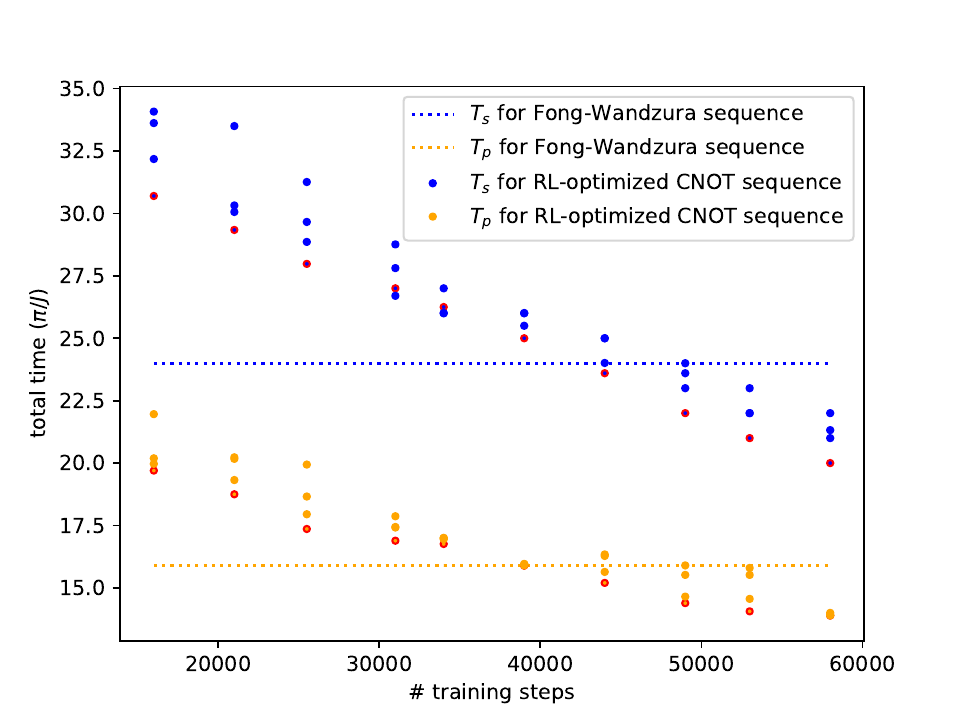}
\caption{Improvement of the total sequential time $T=T_s$ (in blue) and total parallel time $T=T_p$ (in orange) of the CNOT gate sequences discovered by the RL algorithm depending on the number of passed training steps (filled circles). The total gate time $T$ in units of $\pi/J$, where $J$ is the constant strength of the exchange interaction, is plotted as a function of the number of training steps. The solutions with optimal $T_p$ for each investigated number of training steps are highlighted with a red circumference. Note that optimal $T_s$ does not necessarily lead to optimal $T_p$.  For comparison, we show  $T_p$ ($T_s$) of the FW gate sequence 
as an orange (blue) dotted horizontal line.  
The RL procedure is used as described in Sec.~\ref{rlalgo} with a reward  described in Eq.~(\ref{eq:reward}).
We find that after about 45000 training steps, the sequences obtained from the algorithm trained using RL become more time-efficient than known sequences.
}
\label{fig:benchmarks}
\end{figure*}

Multiple exact CNOT gate sequences of 14 time steps were discovered using the RL algorithm for gate sequence optimization. Most of the discovered sequences comprise 14 time steps, but several, including the FW gate sequence and the improved FW gate sequences, required only 13 time steps. The total times  $T_s$ and $T_p$ for performing the exchange pulses sequentially and in parallelized form, respectively, of some of the discovered CNOT gate sequences are plotted as a function of the used training steps in Fig.~\ref{fig:benchmarks}. We find that with an increasing number of training steps, the solution discovered by the algorithm improves. The best solution, discovered by the algorithm has shorter total sequential and parallel times than the original sequence published by Fong and Wandzura in \cite{fong2011universal}.

In addition to the results for the CNOT gate, the RL algorithm also produced several exact CZ sequences of length 14 time steps discovered by the RL algorithm, see Fig.~\ref{fig:CZbenchmark}. 
The dotted lines correspond to the total times required for the parallel and sequential operation of the CZ gate sequence described in \cite{weinstein2023universal}. The shortest sequence has total times $T_p=11.5\,\pi/J$ and $T_s=16.0\,\pi/J$, respectively, for parallel and sequential execution. This sequence (shown in Fig.~\ref{fig:CZ}) is equivalent to the CZ gate described in \cite{weinstein2023universal}. As the number of RL training steps increases, the corresponding best solutions discovered by the algorithm improve in efficiency. 
For details on the results for the CZ gate sequences, see Appendix~\ref{app:resultsCZ}.

 We also use the RL algorithm to optimize the CNOT gate with a different connecting topology. We again discover many exact solutions of a length of seven five-brick blocks, however, the best solution we discover is with a total sequential time $T_s= 20.4\,\pi/J$, and total parallel time $T_p= 15\,\pi/J$.
 Again the efficiency of the discovered gates depends on the number of training steps of the RL algorithm.
  For details on the results for the 2D topology, see Appendix~\ref{app:2D}.

Finally, we also optimize a sequential CNOT gate sequence for the linear arrangement with singlet-triplet pairs at the edges (the ``33" arrangement shown in gray in Fig.~\ref{fig:grid}).
 Under these constraints, we again discover multiple exact CNOT gates of various efficiencies. In this situation, we only evaluate the sequential total time $T_s$. The results are shown in Fig.~\ref{fig:weinsteinbenchmarks} in the Appendix. Importantly, with our RL approach, we rediscover the gate sequence used in \cite{weinstein2023universal}. However, in addition, we discover a few solutions that are more efficient than the sequence in \cite{weinstein2023universal} with respect to total sequential time $T_s$. The best solution is shown in Fig.~\ref{fig:weinstein}.
 This solution is identical to the Weinstein CNOT sequence with respect to the locally-equivalent part but with optimized local gates at the beginning and at the end of the sequence. The efficiency of the discovered CNOT gates again heavily depends on the training steps of the algorithm.
 For more details see Appendix~\ref{app:resultsWeinstein}.

\section{Discussion of the results}
\label{sec:discussion}
For the linear arrangement with singlet-triplet pairs at the inside of the chain (``11" arrangement), our RL approach finds the realization of an exact CNOT gate which improves previously published results regarding the the total gate times $T_s$ and $T_p$.
For the arrangement with singlet-triplet pairs on the outside of a chain (``33" arrangement) as in \cite{weinstein2023universal} we found a sequence with reduced total sequential time $T_s$ compared to the one implemented in \cite{weinstein2023universal}.
These results demonstrate the power of RL applied to the optimization of quantum gates and quantum gate sequences.

We observed that some of the solutions for the CNOT gate sequence found by our RL approach are related to each other by symmetry operations.
Those symmetries are presented in Appendix~\ref{app:theoremsandlemmas} in the form of mathematical lemmas.
Importantly, explicitly implementing these symmetries in the future can boost the performance of the optimization strategy.
These operations themselves are not difficult to understand and are implicitly used already in the literature at least partially, given that what we refer to as an `inverse' FW sequence (or the non-local part of it) is also referred to as `FW sequence' \cite{setiawan2014robust}.
%
%
%
In general, the term `FW sequence' refers to different sequences in the literature \cite{setiawan2014robust,zeuch2020efficient,weinstein2023universal}, see Fig.~\ref{fig:CNOT}, which can be either an explicit CNOT sequence as in the original work by Fong and Wandzura \cite{fong2011universal} or locally equivalent \cite{setiawan2014robust} and which can be either using the same SWAP$^\alpha$ gates presented in \cite{fong2011universal} as in \cite{zeuch2020efficient} or the inverse operations \cite{setiawan2014robust,weinstein2023universal}.
Remarkably, our RL approach found both versions from scratch for the exact CNOT for the same linear arrangement of physical spin qubits as in \cite{fong2011universal}, i.e., the chain highlighted in blue in Fig.~\ref{fig:grid}, see Sec.~\ref{sec:methods} for the details of our optimization procedure.
We note that the CNOT sequence presented here in Fig.~\ref{fig:CNOT} (a) requires a shorter gate time than the original FW sequence \cite{fong2011universal} and in contrast to the one presented in \cite{setiawan2014robust}, it provides the full CNOT sequence rather than a sequence which is locally equivalent to CNOT.
We further note that we did not impose any restrictions on the values of $\alpha$ for the SWAP$^\alpha$ gates in contrast to \cite{setiawan2014robust}.
While a restriction to $\sqrt{\mathrm{SWAP}}$ gates and products as made in \cite{setiawan2014robust} cannot yield an exact CNOT gate, it can, however, provide a gate sequence that is locally equivalent to the CNOT or CZ gate.
The independence from such restrictions on the gate sequence demonstrates the flexibility of our approach.
Regarding the more complex optimization problem for the 2D connectivity, we note that while the RL algorithm can tackle also this problem,
it is challenging to obtain a solution comparable in total time to the most efficient sequence for linear connectivity.

\section{Conclusion}

We have shown that machine learning and intelligent optimization through RL are working approaches for finding optimal exchange-only gate sequences.  Specifically, we have discovered optimized solutions for a variety of gates and connectivities. 
In this work, RL has demonstrated its flexibility and usefulness in optimizing the total times of exchange-only CZ and CNOT gate sequences.
The results demonstrate that RL helps finding such sequences and improves the total gate times of state-of-the-art solutions with fewer prior assumptions compared to other approaches.
In the optimization problems considered in this work, we have used a brick (base) structure that encodes the commutation relations of the exchange coupling. By enforcing a fixed connectivity we have turned the problem into a continuous optimization problem, for which efficient methods exist.
We then use the RL as a tool for intelligent optimization that learns the appropriate starting points of a local optimizer. 
We find optimized solutions that are better or equivalent in terms of total times to known state-of-the-art solutions. 

A limitation of our approach is the use of a fixed brick structure, which captures the commutation relations of operators but is not unique.
Ideally, different brick structures could be used in optimization.
For a more flexible approach, the symmetries that follow from the commutator relations can be encoded in an equivariant neural network, instead of using a fixed brick structure. This is meaningful also for other symmetries in the search space. 
Moreover, symmetries arising from the commutation relationships, as well as the other discovered symmetries, could be exploited by directly incorporating them in various ways in the optimization problem.
As the approach is flexible, it allows for the investigation of different connection topologies in future work. 
Instead of optimizing the total gate time, the objectives could also be to minimize the number of exchange gates or the number of time steps.
This might be particularly promising for gates other than CNOT and CZ.
Additionally, one could extend the problem of optimal exchange-only gate sequences to a more realistic scenario where the gate fidelity is optimized in the e presence of state leakage and noise. 

The RL-based approach presented here is by no means limited to spin exchange-only qubits.
In contrast, it can be broadly applied for finding sequences for various quantum computing hardware platforms or for optimizing compiling sequences of quantum gates.

\section*{Acknowledgements}
We thank Thaddeus D. Ladd for useful correspondence. 
Violeta N. Ivanova-Rohling is funded by the Zukunftskolleg at the University of Konstanz.

\appendix

\section{Reinforcement learning used for optimization}
\label{app:rldetails}
For all optimization problems investigated here, the RL algorithm used is the Deep Deterministic Policy Gradient algorithm (DDPG) \cite{lillicrap2015continuous} which uses the interaction between an actor network and a critic network to learn \cite{konda1999actor} a policy. Moreover, it is a model-free algorithm that does not impose a model or prior knowledge of the world. 

The RL algorithm uses a deterministic policy gradient and can operate over continuous action spaces.
We use the DDPG as implemented in the ``stable baselines-3" package \cite{stable-baselines3}. 
For the definition of the environment, we use the API provided by the package gym in  OpenAI \cite{brockman2016openai}.

We use a gate sequence of a fixed number of 35 parameters, organized into seven blocks of five SWAP$^\alpha$ gates, i.e. bricks, of parameters (physical gate times), similar to the structure FW sequence but without restrictions on the individual exchange pulses. 
However,  an optimal shorter gate sequence length can potentially be reached via this setting. Note that effectively the sequence is shorter if gate times were found to be zero.
Observation, action space, and steps of the algorithm were implemented in OpenAI.
The actor and critic networks used are standard fully connected multilayer perceptron (MLP) networks with three hidden layers, size 64. The following hyperparameter values \cite{smith2018disciplined} are used: an initial learning rate  $\alpha_a = 0.0001$ of the actor, and initial learning rate $\alpha_c = 0.00001$ of the critic networks with a linear learning rate schedule of decrease, and a batch size of 256. A large buffer size of 5000000 allows for a large number of experiences to be stored. 
The observation space, describing the parameter values of the optimization search, is the 35-dimensional space $\left[ 0,2\right]^{35}$.
The action space used is hybrid, which has a continuous component, as well as a discrete component, $\left[-0.4,0.4\right]^{35}\otimes\{0,\ldots,12\}$.
Namely, at each step, the agent performs an action that is a change of the normalized times $\alpha_i, i=0,\ldots,34$,  as well as    
 a partial Powell optimization is performed with a possible number of iterations from $0$ to $12$, where $0$ iterations means that no Powell derivative-free optimization is performed. The goal is for the algorithm to learn, starting from a  random point, an appropriate sequence of parameter changes for the respective $\{\alpha_0,\ldots\alpha_{34}\}$ as well as appropriate Powell iteration values to navigate the search space. 
 
\section{Results for the CZ-gate}
\label{app:resultsCZ}
In order to adapt the algorithm for the discovery of the CZ gate, all we need is to redefine the reward function. The five-brick structure and the parametrization can be reused directly as defined in the search for optimal CNOT gate sequence. This demonstrates the flexibility of the approach. The quality measure, which evaluates a gate sequence's distance from the CZ gate is given by:
\begin{equation}
\begin{split}
    d_{CZ} = &\left[2-\frac{1}{4}
    \left|U^{(0)}_{11} + U^{(0)}_{22} + U^{(0)}_{33} - U^{(0)}_{44}\right|\right.\\
    &\left.-\frac{1}{4}\left|U^{(1)}_{11} + U^{(1)}_{22} + U^{(1)}_{33} - U^{(1)}_{44}\right| \right]^{1/2}.
\end{split}
\end{equation}
Accordingly, the reward is  given by
\begin{equation}
    R(\boldsymbol\alpha) = N -d_{CZ}(\boldsymbol\alpha) -\gamma T.
    \label{eq:rewardCZ}
\end{equation}
In the following, we present the results found by our RL approach for an exact CZ gate, using the objective function from Eq.~(\ref{eq:rewardCZ}).
\label{sec:resultsCZ}
\begin{figure}
\begin{center}
\includegraphics[width=\columnwidth]{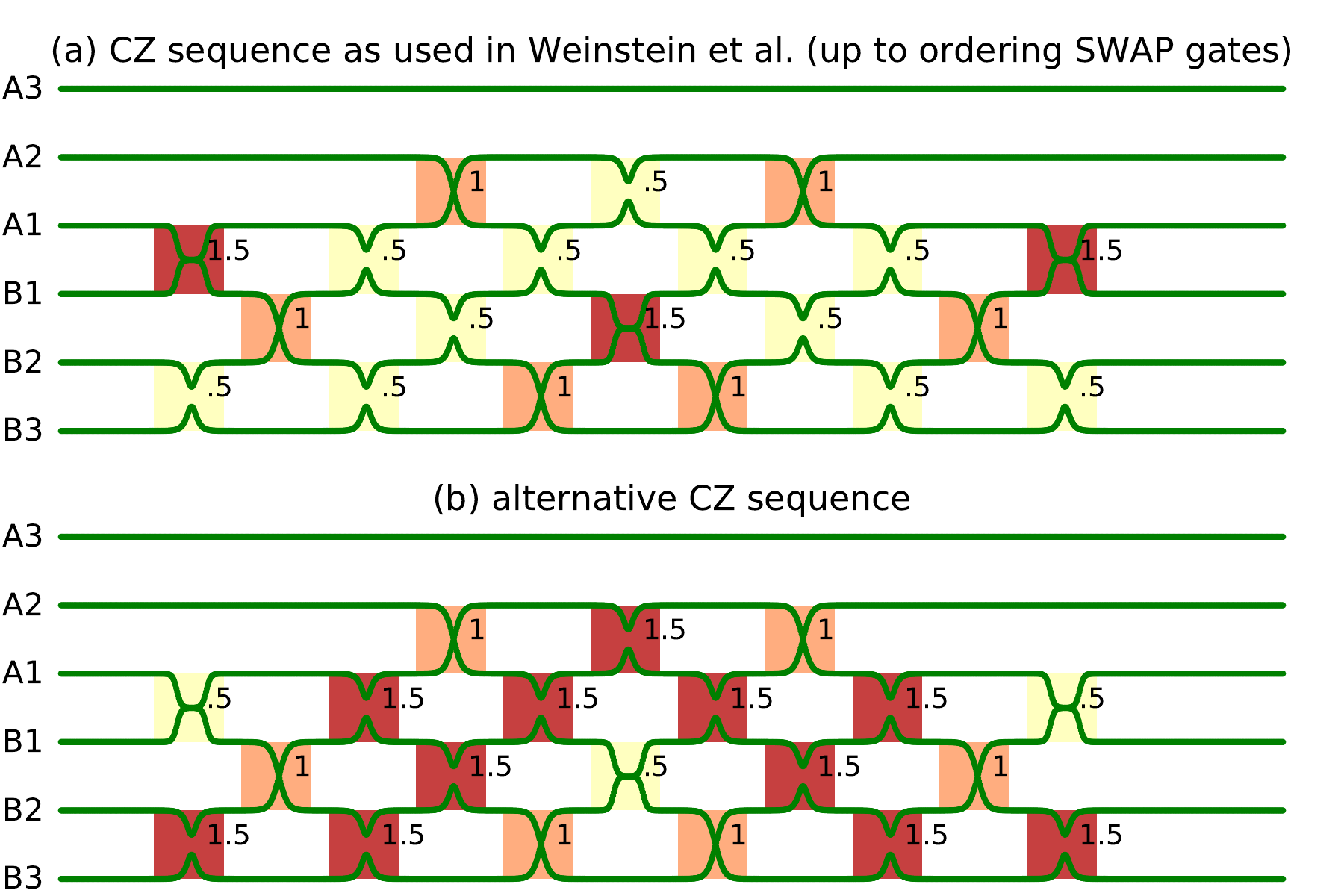}
\end{center}
\caption{(a) The optimal CZ sequence rediscovered by the RL algorithm equivalent to the CZ gate sequence in Weinstein \textit{et al.} \cite{weinstein2023universal} (up to spin-order related SWAP gates) and up to local gates to the sequence in \cite{setiawan2014robust}.
(b) Alternative CZ gate sequence also found by our RL approach and related to the original FW sequence \cite{fong2011universal} by altering the local operations on logical qubits A and B at the beginning and end of the sequence.
The sequences shown in (a) and (b) are related to each other by 'inverting' the sequence, see Appendix \ref{app:theoremsandlemmas}. They are equivalent in number of exchange pulses and number of time steps when using parallelism. Sequence (a) is superior in both, $T_p$ and $T_s$.}
\label{fig:CZ}
\end{figure}
The results obtained are similar to the sequences found for CNOT, see Fig.~\ref{fig:CNOT}.
Again there are solutions which are related to each other by symmetry transformations, see Appendix~\ref{app:theoremsandlemmas}.
The sequence in Fig.~\ref{fig:CZ} (b) is related to the original FW sequence \cite{fong2011universal} by changes only in the local gates for the A and for the B qubit at the beginning and at the end of the sequence.
On the other hand, the sequence in Fig.~\ref{fig:CZ} (a) is up to reordering of the spins-related SWAP gates identical to the CZ sequence from \cite{weinstein2023universal} where it is referred to as `FW CZ' sequence.

\begin{figure*}
\includegraphics[width=\textwidth]{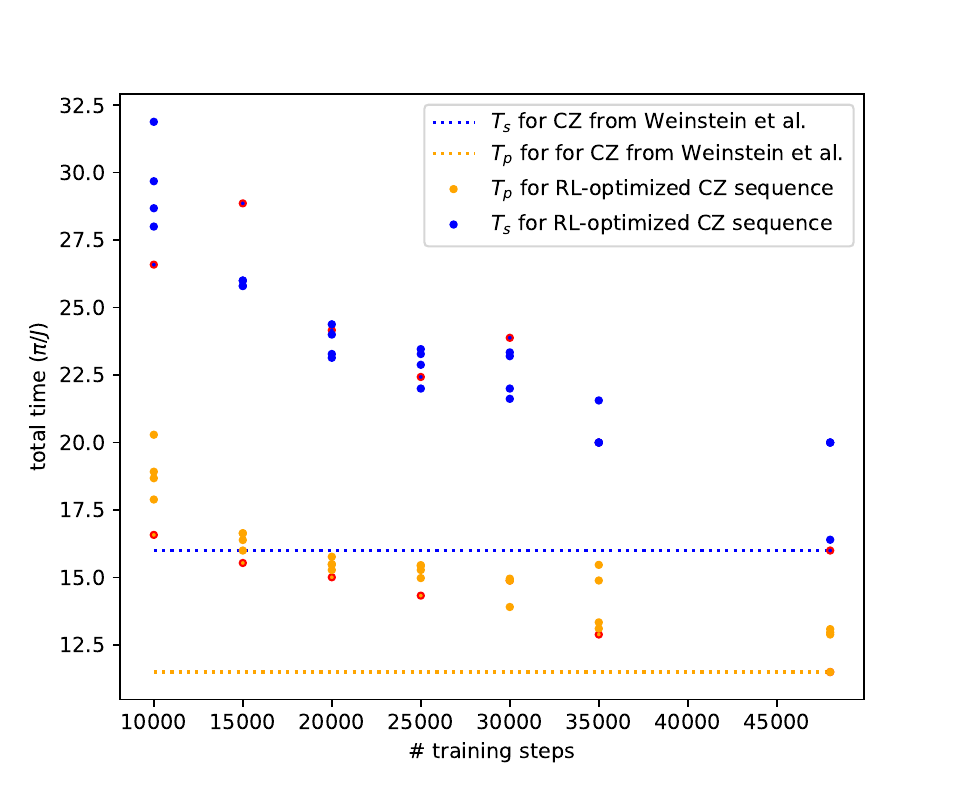}
\caption{Improvement of $T_s$ (in blue) and $T_p$ (in orange)  of the CZ gate sequences units of $\pi/J$ discovered by the RL algorithm depending on the number of training steps used.  Further details, see Fig.~\ref{fig:benchmarks}.}
\label{fig:CZbenchmark}
\end{figure*}

The performance of the RL algorithm as a function of training steps is presented in Fig.~\ref{fig:CZbenchmark}.
While the CNOT or CZ sequences can be generated from each other by padding with local single-qubit gates acting on the qubits A and B at the beginning and at the end of the sequence, the fact that our RL approach performs well for finding both, CNOT and CZ sequences from scratch is an important indication for the flexibility of the ansatz.

\section{Optimizing gate sequences with fixed order gates to compare to the solution presented in \cite{weinstein2023universal}}
\label{app:resultsWeinstein}
\begin{figure}
\includegraphics[width=9cm]{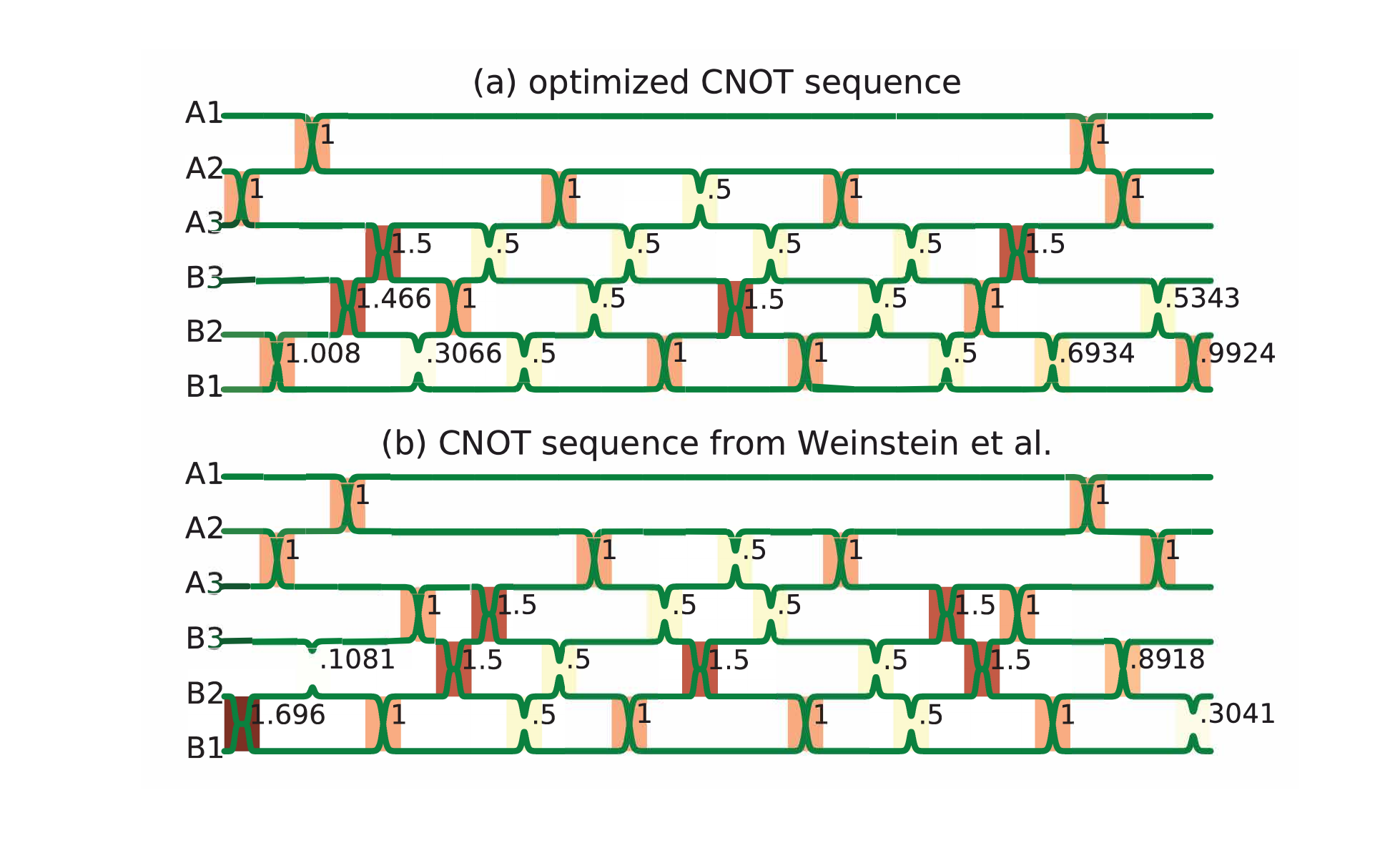}
\caption{CNOT gate sequences for the linear arrangement with the singlet-triplet pairs on the outside (highlighted in gray in Fig.~\ref{fig:grid}): The RL-optimized sequence of sequential time $T_s=24\pi/J$ and the sequence from \cite{weinstein2023universal} of $T_s=26\pi/J$.}
\label{fig:weinstein}
\end{figure}

The arrangement in Ref.~\cite{weinstein2023universal} is such that the singlet-triplet pairs are on the ends of the chain of six spin qubits, as in the chain highlighted in gray in Fig.~\ref{fig:grid}.
This explains that the sequence implemented by Weinstein \textit{et al.} has eight additional  SWAP gates which are in some sense switching between two distinct linear orders (highlighted in blue and gray in Fig.~\ref{fig:grid}).
The total sequential time of the Weinstein CNOT sequence, see Fig.~\ref{fig:weinstein}, is $26 \pi/J$ assuming a constant exchange coupling $J$ for each of the exchange gates in contrast to the actual implementation in \cite{weinstein2023universal}, while the total time of the improved FW sequence together with eight ordering SWAP gates is $28 \pi/J$.
However, this is an unfair comparison, as some gates at the beginning and the end are shifted relative to the ordering SWAP gates to transform them in a more efficient way.
In order to use the Weinstein CNOT sequence as a benchmark, we need to use our RL approach applied to their architecture and then minimize the sequential time.


To be able to fairly compare to the sequence discussed in \cite{weinstein2023universal}, we set specific constraints for the gate sequence -- sequential execution of the physical gates and we use the same pairs spins coupled by exchange gates as in \cite{weinstein2023universal}. RL was used to optimize a CNOT sequence that is sequential and with order gates using the form from \cite{weinstein2023universal}. We achieved sequences with better total sequential time than the one in \cite{weinstein2023universal}, see Fig.~\ref{fig:weinstein}.
 \begin{figure*}
\includegraphics[width=12cm]{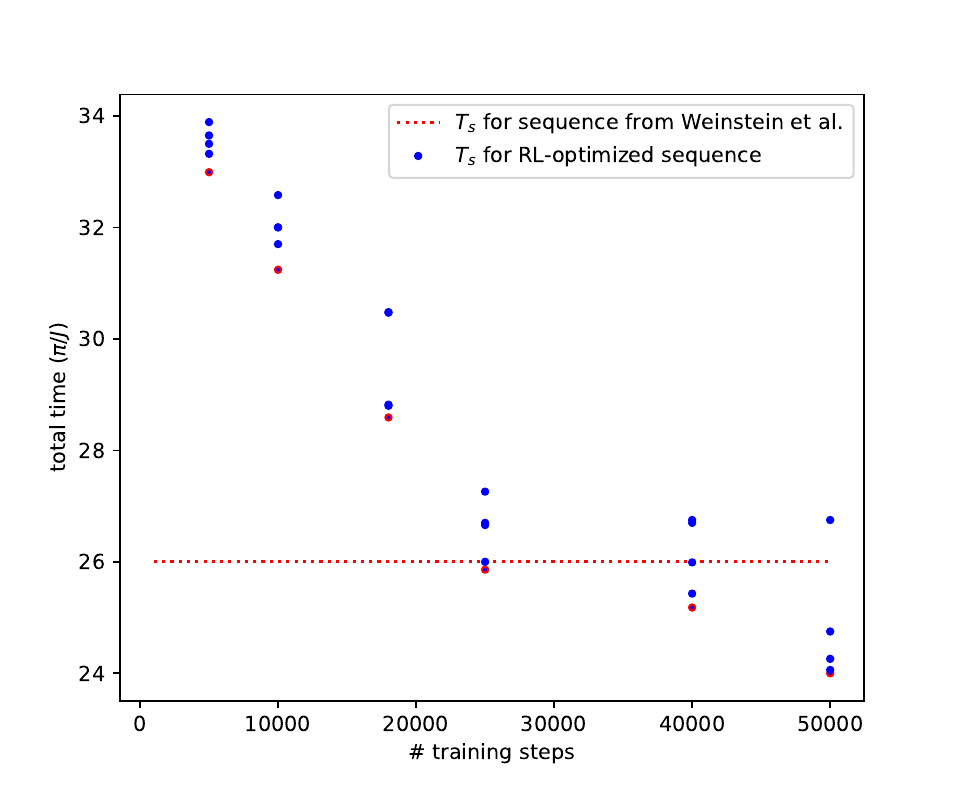}
\caption{Total times of the resulting CNOT gate sequences with fixed ordered gates discovered by the RL algorithm and the sufficient number of training steps, needed to obtain them.}
\label{fig:weinsteinbenchmarks}
\end{figure*}

\section{2D architecture optimization}
\label{app:2D}
In addition, we apply RL to optimize a two-dimensional (2D) topology where each spin is exchange-coupled to a spin of the other qubit.
\begin{figure}
\includegraphics[width=\columnwidth]{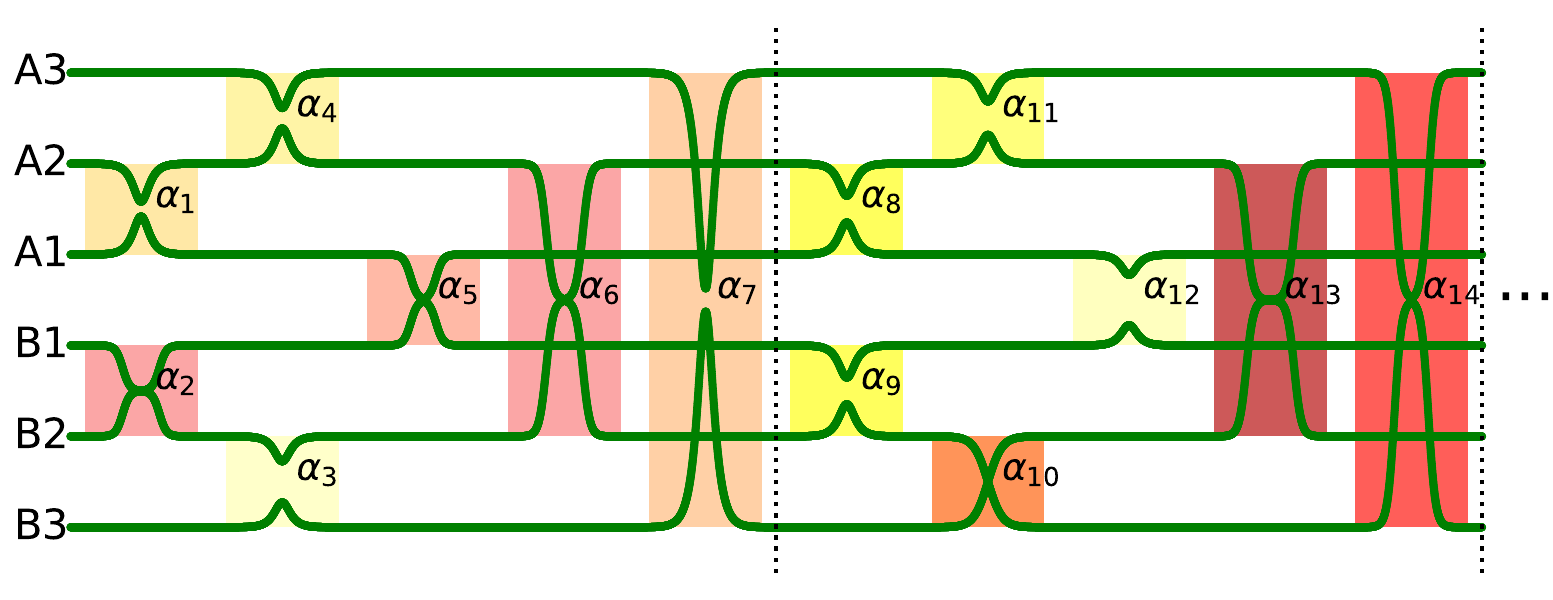}
\caption{The seven-parameter block of quantum gates used for the physical qubits being arranged in a two-by-three rectangle with nearest-neighbor coupling. Note that the exchange gates between two spins where one belongs to logical qubit A and the other to logical qubit B (corresponding normalized gate times are $\alpha_5$, $\alpha_6$, $\alpha_7$ or $\alpha_{12}$, $\alpha_{13}$, $\alpha_{14}$) can be performed in parallel.}
\label{fig:7Block}
\end{figure}
The constraints of the 2D arrangement, highlighted in green in Fig.~\ref{fig:grid}, lead to a modification of the 5-brick structure used in the FW optimization, yielding a seven-component structure as shown in Fig~\ref{fig:7Block}. In each block, we first apply the exchange gates between the logical qubits A1 and A2 as well as B1 and B2 in parallel. Second, we apply the gates between A2 and A3 as well as between B2 and B3 in parallel. Finally, we apply the interactions $J_{Aj Bj}$ with $j=1,2,3$ in parallel. We find a  gate sequence for CNOT with total time $T_s=20\,\pi/J$ ($T_p=15.89\,\pi/J$).


\section{Theoretical derivation of observed symmetries}
\label{app:theoremsandlemmas}
Among the solutions obtained by RL, we observe that some are related to each other by a symmetry operation that leaves the resulting logical quantum gate unaffected.
We elaborate on such symmetry operations in the following, using some specific properties of the CNOT and CZ gates, namely that these unitaries are also Hermitian and that -- in matrix form -- they have real entries.

We start by reminding ourselves that switching the sign of the time a constant Hamiltonian is "switched on" translates from a unitary operation to its inverse:
\begin{lemma} A quantum gate given by
$U(t) = \exp\left(\frac{-it}{\hbar} H\right)$
with a time-independent Hamiltonian $H$, fulfills
$U^{-1}(t) = U(-t).$
\end{lemma}
In the context described in this paper, unitary operations are given by a sequence of consecutive gates of the form given in the lemma above.
This means constant Hamiltonians $H_j$ are switched on for times $t_j$.
We directly obtain the following statement about those sequences.
\begin{lemma}
    For a gate sequence $U(t_1,\ldots,t_n) = U_n(t_n)\cdots U_1(t_1)$ with $U_j(t_j)= \exp\left(\frac{-it_j}{\hbar} H_j\right)$, the inverse unitary operation is given by the sequence $[U(t_1,\ldots,t_n)]^{-1} = U_1(-t_1)\cdots U_n(-t_n).$
\end{lemma}
Consequently, we can express a unitary operation which is Hermitian, like $\mathrm{CNOT}$:
\begin{lemma}
\label{th:inverse}
    For a gate sequence $U(t_1,\ldots,t_n)$, defined in the same way as in the lemma above with $U^\dag=U$, the reverse sequence with inverted time arguments represents the same unitary,
    $$U_n(t_n)\cdots U_1(t_1) = U_1(-t_1)\cdots U_n(-t_n).$$
\end{lemma}
For sequences of palindromic structure, this yields:
\begin{corollary}
\label{coro:palin}
    If a gate sequence is given by $U(t_1,\ldots,t_k) = U_1(t_1)\cdots U_k(t_k)\cdots U_1(t_1)$, i.e., it is of palindromic structure, and obeys $U^\dag = U$, then the same unitary operation is represented by the sequence with negative time arguments,
    $$U(t_1,\ldots,t_k) = U(-t_1,\ldots,-t_k).$$
\end{corollary}
Note that the non-local part of the FW sequence for $\mathrm{CNOT}$ and also an exact CZ sequence are palindromic.

Now, we will use the properties of the distance measures $d_{\rm FW}$ and $d_{\rm CZ}$ and properties of the individual exchange gates.
These individual gates for the exchange-only qubits are $\mathrm{SWAP}^\alpha$ gates applied to two of the physical qubits.
A $\mathrm{SWAP}^\alpha$ gate is symmetric in the standard basis $\{|00\rangle,|01\rangle,|10\rangle,|11\rangle\}$ as well as in the representation used for computing the exchange-only sequences for the logical subsystem ($5\times5$ for zero spin, $9\times9$ matrices for spin one), see  Ref.~\cite{fong2011universal}.
From this it follows that a sequence of $\mathrm{SWAP}^\alpha$ gates, $U(\alpha_0,\ldots,\alpha_{n-1}) = \mathrm{SWAP}_{i_nj_n}^{\alpha_n} \cdots \mathrm{SWAP}_{i_1j_1}^{\alpha_1}$ fulfills
$$U(2-\alpha_1,\ldots,2-\alpha_n)= [U(\alpha_1,\ldots,\alpha_n)]^*,$$
where $*$ denotes the complex conjugate (not the Hermitian conjugate).


Note that CNOT has only real entries in the logical subsystem in the standard basis.
Consequently, two gates which are complex conjugate to each other, have the same distance to CNOT.
This holds for the Euclidean distance as well as for the FW loss function, $d_{\rm FW}$.

\begin{lemma}
\label{th:distance}
    For sequences of $\mathrm{SWAP}_{ij}^\alpha$ gates, 
    $$ U(\alpha_1,\ldots,\alpha_n) = \mathrm{SWAP}_{i_nj_n}^{\alpha_n} \cdots \mathrm{SWAP}_{i_1j_1}^{\alpha_1},$$ the following holds
    $$d_{\rm FW}(U(\alpha_1,\ldots,\alpha_n)) = d_{\rm FW}(U(2-\alpha_1,\ldots,2-\alpha_n)).$$ 
\end{lemma}

Note that the lemma above also holds for $d_{\rm CZ}$.
Further note that Lemma~\ref{th:inverse} and Lemma~\ref{th:distance} applied to a CNOT (CZ) sequence yield two sequences also representing CNOT (CZ).
Only if the sequence is of palindromic structure these two sequences are identical to each other, compare Corollary \ref{coro:palin}.



\bibliography{bibliographyEORL}
\end{document}